\documentclass[twocolumn,secnumarabic,amssymb, aps,prl,reprint]{revtex4-1}
\usepackage{graphicx}
\usepackage{color}
\usepackage{dcolumn}
\usepackage{bm}
\usepackage{ucs}
\usepackage{physics}
\usepackage{mathtools}
\begin{document}

\title{Supercurrent in a double quantum dot}

\author{J. C. Estrada Salda\~{n}a$^{1}$}
\author{A. Vekris$^{1}$}
\author{G. Steffensen$^{1}$}
\author{R. \v{Z}itko$^{2}$}
\author{P. Krogstrup$^{1}$}
\author{J. Paaske$^{1}$}
\author{K. Grove-Rasmussen$^{1}$}
\author{J. Nyg{\aa}rd$^{1}$}

\email{Corresponding author: nygard@nbi.ku.dk}

\affiliation{$^{1}$Center for Quantum Devices, Niels Bohr Institute, University of Copenhagen, 2100 Copenhagen, Denmark}
\affiliation{$^{2}$Jo\v{z}ef Stefan Institute, Jamova 39, Ljubljana, Slovenia \\
Faculty of Mathematics and Physics, University of Ljubljana, Jadranska 19, Ljubljana, Slovenia}

\date{\today}

\begin{abstract}
We demonstrate the Josephson effect in a serial double quantum dot defined in a nanowire with epitaxial superconducting leads. The supercurrent stability diagram adopts a honeycomb pattern with electron-hole and left-right reflection symmetry. We observe sharp discontinuities in the magnitude of the critical current, $I_c$, as a function of dot occupation, related to doublet to singlet ground state transitions. Detuning of the energy levels offers a tuning knob for $I_c$, which attains a maximum at zero detuning. The consistency between experiment and theory indicates that our device is a faithful realization of the two-impurity Anderson model.
\end{abstract}

\maketitle
Recent progress in the microfabrication of hybrid semiconductor-superconductor one-dimensional heterostructures has resulted in Josephson junctions with controlled ground states \cite{de2010hybrid}. This control is achieved by confining electrons in a quantum dot (QD) in the semiconductor weak link.
In a weakly-coupled QD, the number of electrons, $n$, is known precisely thanks to the charging energy, $U$, and it is conveniently controlled by a gate voltage.  
A singly-occupied dot shifts the phase difference between the superconducting leads by $\pi$, which corresponds to a doublet ground state.  
When $n$ changes in an even-odd-even fashion, an alternating $0-\pi-0$  pattern in the phase of the supercurrent is expected. This has been experimentally observed in single QDs \cite{van2006supercurrent,cleuziou2006carbon, Jorgensen2007,Eichler2009,Maurand2012,delagrange2015manipulating,delagrange2017}.

A serial double quantum dot (DQD) Josephson junction offers greater freedom in the control of the ground state of the system. DQD levels, as opposed to multi-level single dots \cite{van2006supercurrent,shimizu1998multilevel,lee2010josephson,karrasch2011supercurrent}, have the advantage of individual gate control of the parameters of each dot level. The subgap states of a DQD attached to two superconducting leads have been previously investigated, but the soft gap of the device resulted in an effectively S-DQD-N system (N stands for a normal metal, and S for a superconducting lead), and a supercurrent has not been reported \cite{su2017andreev}. Subgap states have also been described in a purposely engineered S-DQD-N device, in a regime with stronger coupling to the S lead, which could then screen the spin of the dots \cite{grove2017yu}. Chains of  dots in series coupled by superconductors can be mapped to the Hamiltonian of a topological superconductor, which constitutes an additional motivation for investigating this type of devices \cite{sau2012realizing,fulga2013adaptive}.

\begin{figure} [htb!]
\includegraphics[width=1\linewidth]{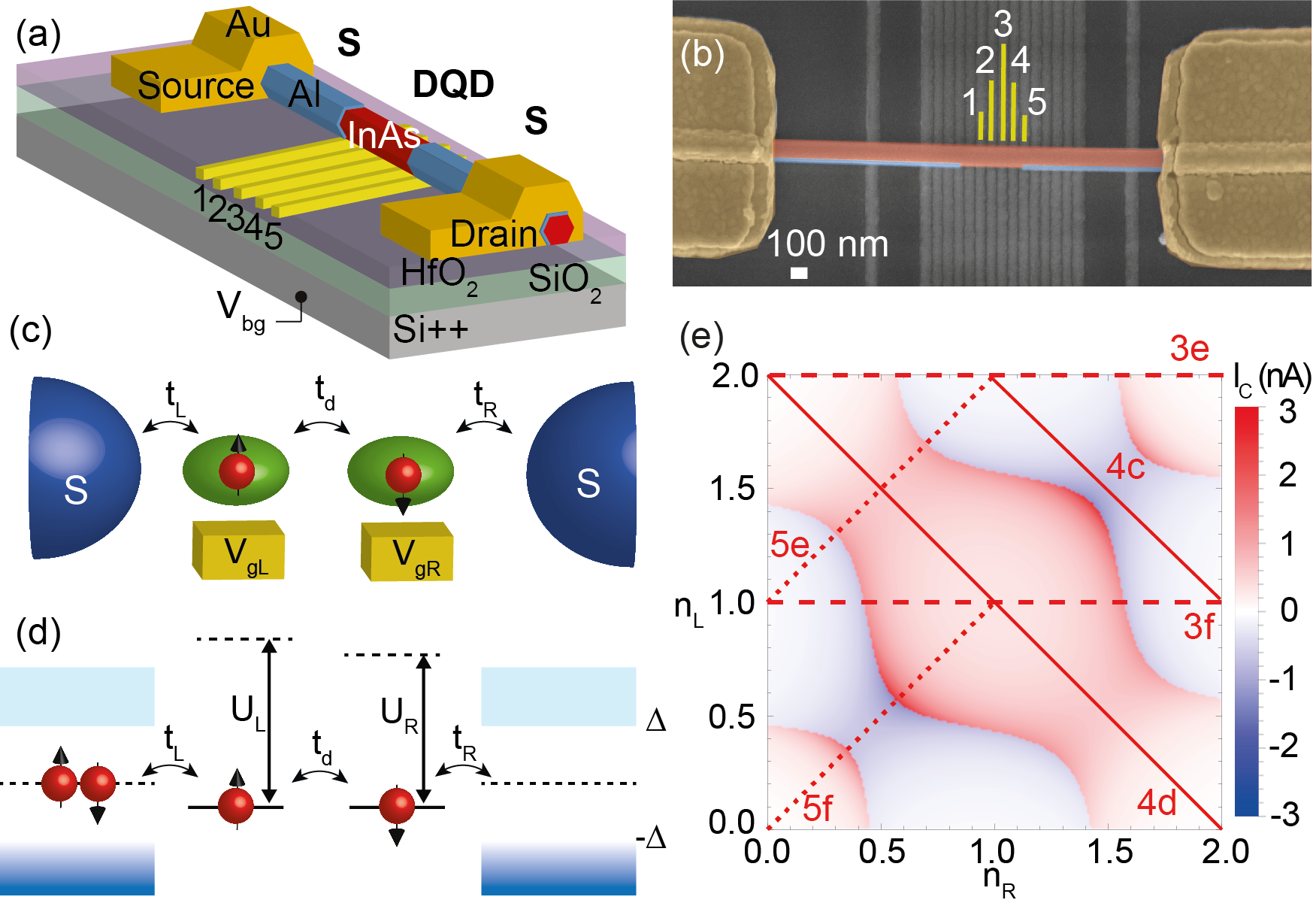}
\caption{(a) Schematics of the device. Gates numbered 1 to 5 were used to electrostatically define two quantum dots in the channel of the bare InAs nanowire. A backgate voltage, $V_{bg}$, was used for further fine-tuning of the coupling of the dots. (b) Scanning-electron micrograph of the measured device. Au contacts, epitaxial Al leads, and InAs nanowire are colored in yellow, blue and red, respectively. (c) Schematics of two quantum dots in series, whose energy level is tuned by the voltage on gates $V_{gL}$ and $V_{gR}$ (for left and right dot), which in our device correspond to gates 2 and 4. (d) Energy diagram of the system, consisting of two single levels in series with energy levels $\epsilon_L$ and $\epsilon_R$, coupled to the closest superconducting lead with tunnel couplings $t_L$ and $t_R$, respectively, and with inter-dot coupling $t_d$. The charging energies, $U_L \approx 1.85$ meV and $U_R \approx 1.62$ meV, are much larger than the superconducting gap, $\Delta=0.265$ meV. The large level spacings, $\Delta E_L\approx 1.64$ meV and $\Delta E_R\approx 1.76$ meV, of the order of $U_L$ and $U_R$, respectively, make the single-level approximation valid. (e) Charge stability diagram $I_c(n_L,n_R)$ calculated through fourth order perturbation theory \cite{NoteParameters}.}
\label{fig1}
\end{figure}

In this work, we demonstrate for the first time a gate-controlled S-DQD-S Josephson junction, and focus on a regime of weak coupling to the two S leads. We find discontinuities in the critical current which depend on the total number of electrons in the dots. Their lineshape, characterized by asymmetries and changes in magnitude, allows us to identify them as $0 \to \pi$ phase transitions corresponding to singlet to doublet ground state changes, as previously done for single QDs \cite{van2006supercurrent,cleuziou2006carbon, Jorgensen2007, Maurand2012,delagrange2017,estrada2018supercurrent}. Whenever an odd (even) total number of electrons is present in the highest-lying levels of each dot, the DQD ground state is a doublet (singlet). Additionally, we find a modulation of $I_c$ through orbital detuning, with maximal $I_c$ at maximal orbital hybridization. The changes in the ground state are corroborated by crossings of subgap states. We model our findings using the two-impurity serial Anderson model.

\begin{figure} [htb!]
\includegraphics[width=1\linewidth]{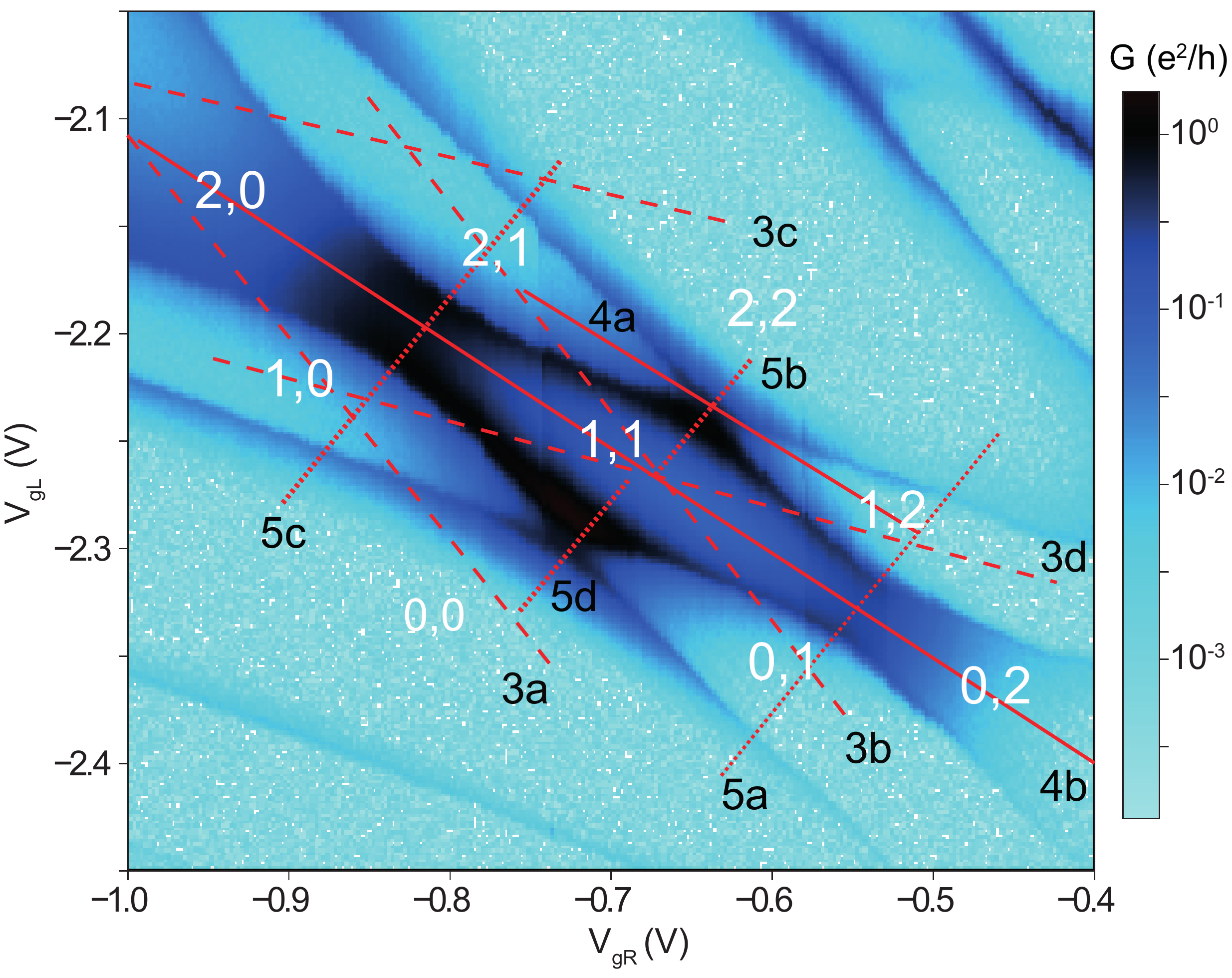}
\caption{Charge stability diagram as a function of plunger gate voltages. Effective charge numbers $n_L,n_R$ determined from the shell-filling pattern (see Supplemental Material) are indicated for each sector. Lines on the diagram indicate line-cuts shown in subsequent figures. $T=15$ mK, $V_{sd}=0$, $V_{g1}= -9.2$ V, $V_{g3}= -9.3$ V, $V_{g5}= -0.25$ V, and $V_{bg}= 10.4$ V.}
\label{fig2}
\end{figure}

\begin{figure} [htb!]
\includegraphics[width=1\linewidth]{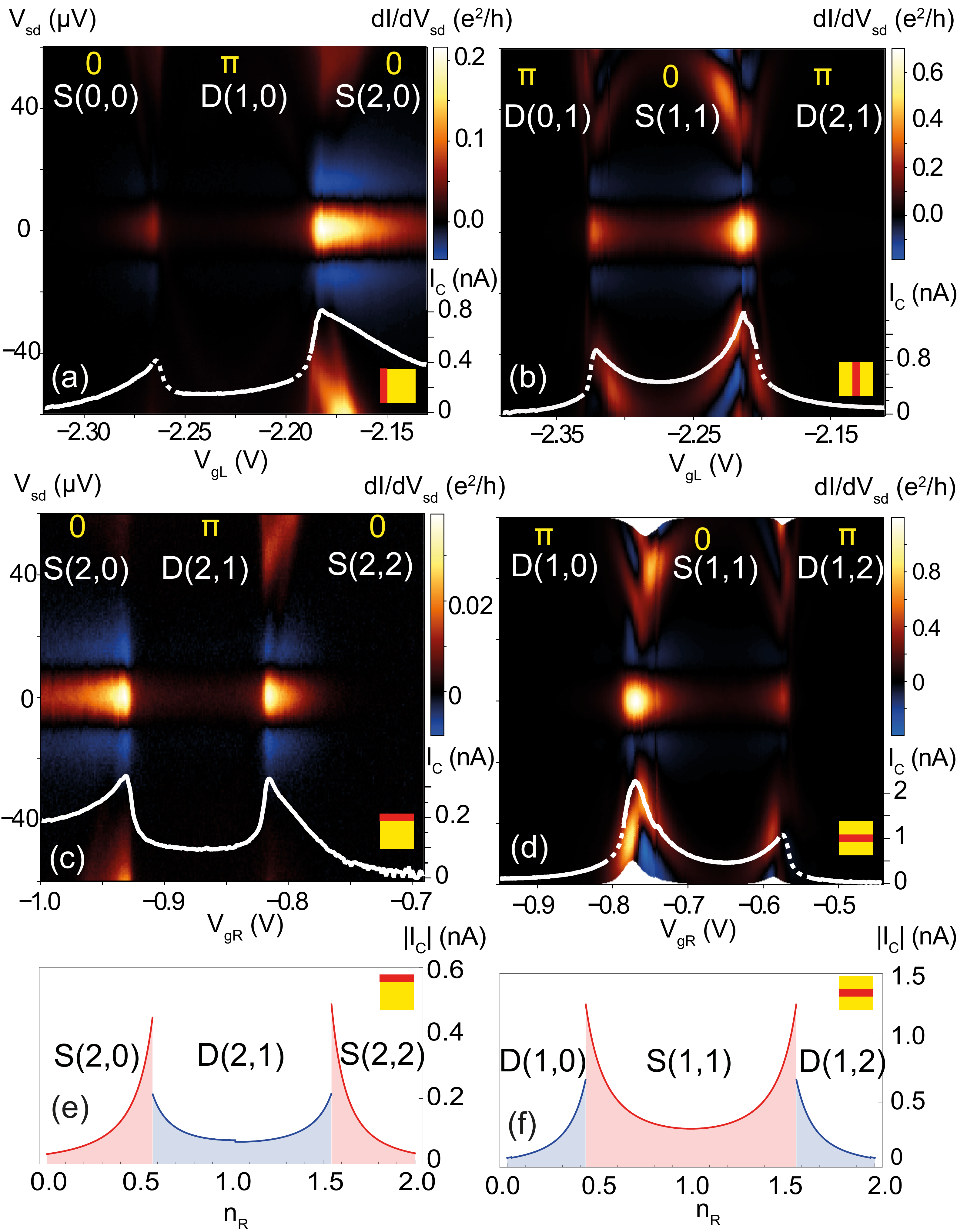}
\caption{(a-d) Line-cuts along the dashed lines in Fig.~\ref{fig2} for fixed $n_R$ (a,b) and for fixed $n_L$ (c,d). \textit{Overlays}. Fitted $I_c$ (white curve) \cite{NoteFit}. The notation used to name each sector is GS$(n_L,n_R)$, where GS=D, S represents the ground state of the sector. (e,f) Line-cuts of $|I_c|$ calculated by fourth order perturbation theory along trajectories in occupation number (proportional to gate voltage) given in Fig.~\ref{fig1}e. Red (blue) color indicates that $I_c$ is positive (negative). Line-cut trajectories in Figs.~\ref{fig1}e and \ref{fig2} are symbolically indicated on each panel.}
\label{fig3}
\end{figure}

Our device is depicted in Figs.~\ref{fig1}a,b. A 110 nm-diameter InAs nanowire with an in-situ grown epitaxial 7 nm Al shell \cite{krogstrup2015epitaxy}, covering three of its facets, was first deterministically placed over a bed of finger gates insulated by a 20 nm layer of hafnium oxide dielectric. By selectively etching the Al shell over five of these gates, we obtained a 350 nm section of bare InAs nanowire. Afterwards, we contacted the ends of the wire with two Ti/Au strips, each at a distance roughly 1 $\mu$m away from the bare wire. Additionally, we used as a global backgate the highly-doped Si substrate with a 300 nm Si oxide on which the device was fabricated.

Two QDs in series were defined in the bare wire through gate tuning, and were proximitized by the sections of the wire with superconducting Al. A scheme of this S-DQD-S system and its energetics are portrayed in Figs.~\ref{fig1}c,d, respectively. All differential conductance measurements were done in a dilution cryostat at 15 mK, using a standard lock-in technique with an AC excitation of 2 $\mu$V at a frequency of 116.81 Hz.  

Figure \ref{fig2} shows a colormap of the zero-bias linear conductance ($G$) of the device as a function of the left and right plunger gates, $V_{gL}$ and $V_{gR}$, reminiscent of the \textit{honeycomb} charge stability diagram of normal-state DQDs \cite{van2002electron}. This similarity stems from parity changes nearly matching charge degeneracies in the low-coupling regime. For clarity, only the number of electrons $n_L$=0,1,2 and $n_R$=0,1,2 in the highest-lying levels of each dot, left and right, respectively, are shown, as identified by the apparent shell filling pattern \cite{jorgensen2008singlet} (see Supplemental Material). Though this measurement is analogous to the stability diagram of a conventional N-DQD-N system, it is important to note that in our Josephson DQD circuit $G$ originates from a supercurrent. Cotunneling processes which contribute to this supercurrent are more probable at charge degeneracies, and are particularly enhanced at the quadruple points where the charge states ($n_L$,$n_R$), ($n_L+1$,$n_R$), ($n_L$,$n_R+1$) and ($n_L+1$,$n_R+1$) are degenerate in the case $U_{LR}=t_d=0$, leading to an increase of $G$ near these points. The diagram is electron-hole symmetric and, due to similar $U_L$ and $U_R$, it is also left-right symmetric. Nonetheless, there is a slight asymmetry in the conductance with respect to the diagonal from $(0,0)$ to $(2,2)$, due to cross coupling of $V_{gL}$ with the left tunnel barrier, which is also evidenced by the lower $G$ of charge sectors at more negative $V_{gL}$ (see Supplemental Material). Fig.~\ref{fig1}e shows the calculated critical current, $I_c$, for the parameters of the device and fixed tunnel barrier \cite{NoteParameters}, using fourth order perturbation theory in the tunnel coupling (for additional and qualitatively consistent NRG and zero-bandwidth calculations, and details of the theory, see Supplemental Material). These calculations confirm the experimental observation that $I_c$ is largest close to degeneracy points and is especially enhanced near the quadruple points. The particle-hole symmetry of the underlying Hamiltonian is manifested in the figure through the relation $I_c(n_L,n_R)=I_c(2-n_L,2-n_R)$. The sign change of $I_c$ going from singlet to doublet sectors indicates the $0-\pi$ transition. 

\begin{figure} [htb!]
\includegraphics[width=1\linewidth]{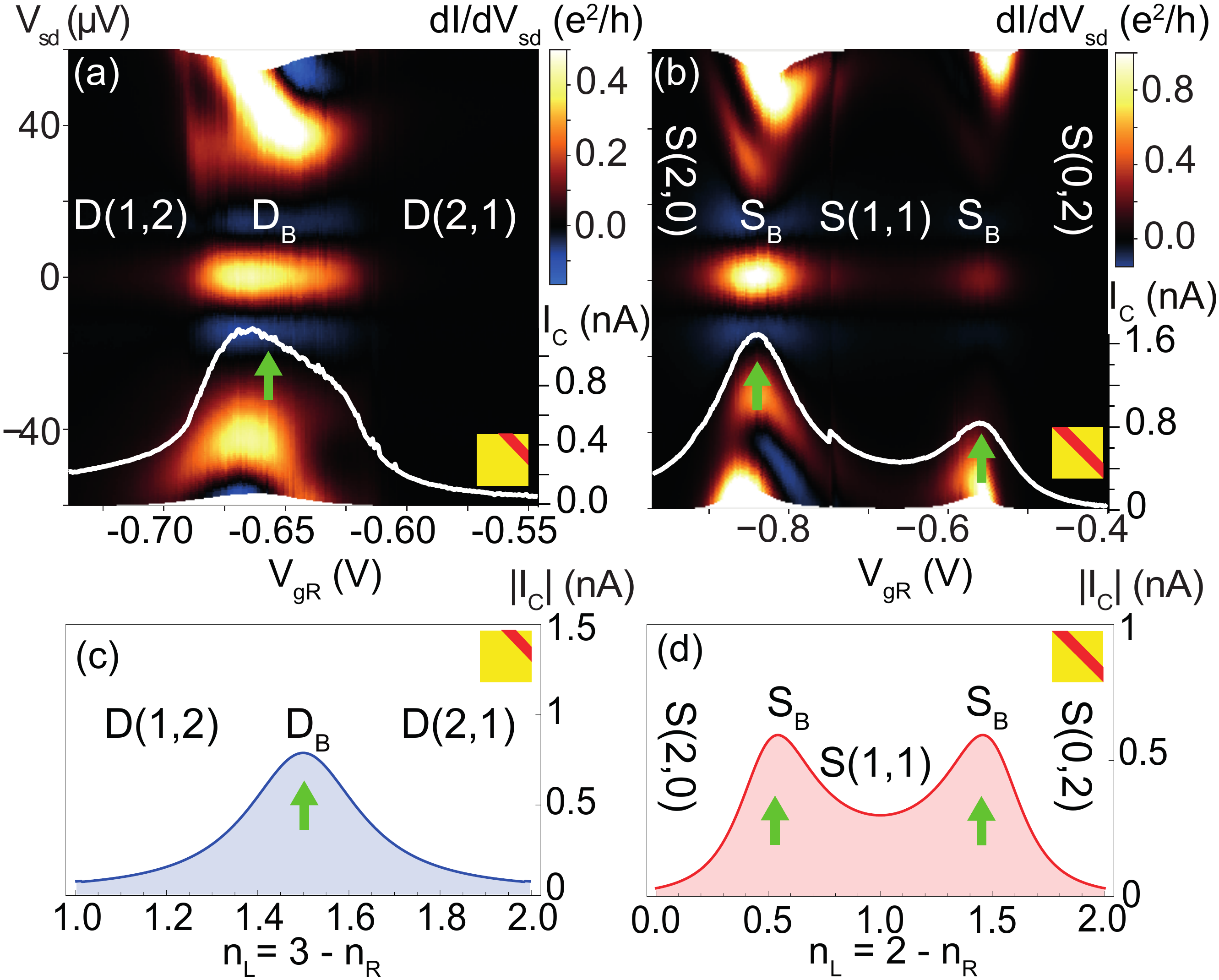}
\caption{(a,b) Detuning $\epsilon_L-\epsilon_R$ line-cuts along the continuous lines in Fig. \ref{fig2}. \textit{Overlays}. Fitted $I_c$ (white line). Note that the ground state, either doublet or singlet, does not change in these line-cuts. (c,d) $|I_c|$ traces calculated by fourth order perturbation theory. Red (blue) accounts for positive (negative) $I_c$. Green arrows indicate zero-detuning points.}
\label{fig5}
\end{figure}

In the following, we present maps of differential conductance, $dI/dV_{sd}(V_{sd})$, zoomed at low-bias within the superconducting gap, along the cuts indicated in Fig.~\ref{fig2}. In Figs.~\ref{fig3}a,b, we show cuts taken at fixed $n_R$ while varying $n_L$, corresponding to a variation of the left-dot energy level. In turn, cuts taken at fixed $n_L$ while varying $n_R$ are shown in panels (c,d). Notably, a narrow zero-bias peak surrounded by negative differential conductance (NDC) traverses all four diagrams. Additional features seen here at $V_{sd}>20$ $\mu$V, also observed at larger bias between $\pm 2\Delta = 0.53$ mV, are discussed in Supplemental Material.

The zero-bias peak in Figs.~\ref{fig3}a-d has a peculiar gate dependence, exhibiting a noticeable discontinuity in the conductance value at every parity change. It sharply drops whenever $n_L+n_R$, the total charge in the dots, is an odd number, with respect to its value at an even number.

We relate this zero-bias peak to a dissipative supercurrent. NDC comes from switching to low subgap current (see Supplemental Material for an $I-V_{sd}$ curve evidencing this). Dissipation can be ascribed within the resistively and capacitively shunted junction (RCSJ) model to the overdamped nature of the junction and to the presence of thermal noise \cite{Anchenko1969,steinbach2001direct,Jorgensen2007,Eichler2009,jorgensen2009critical,estrada2018supercurrent}. By fitting $dI/dV_{sd}(V_{sd})$ traces of Figs.~\ref{fig3}a-d to this model, we extract $I_c$ as a function of the gate voltage (for details of the fit and a justification of the overdamped nature of the junction, see the Supplemental Material) and plot it as overlays in the respective colormaps (white curves).

The observed discontinuities in the conductance of the zero-bias peak are seen as abrupt jumps in the magnitude of the supercurrent, with peaked values at parity changes. We interpret these as 0 $\to \pi$ transitions in the phase-shift $\phi_0$ of the Josephson current $I=I_c\sin(\phi+\phi_0)$. These happen whenever $n_L+n_R$ changes from an even number ($\phi_0$=0) to an odd number ($\phi_0=\pi$). In this regime, in which both dots are relatively decoupled from the superconducting leads, the following rule of thumb applies. When $n_L+n_R$ is even, the ground state of the DQD is a spin singlet, S, whereas when $n_L+n_R$ is odd, it changes to a doublet, D. We are able to distinguish $\pi$ from 0 regions due to the small magnitude of the supercurrent within the $\pi$ regions and due to the asymmetry of the $I_c$ peaks at the parity changes. This asymmetry finds its origin in the halving of cotunneling processes in the $\pi$ domain with respect to the $0$ domain \cite{glazman1989resonant}. An example of these two observations is seen in Fig.~\ref{fig3}c, where $I_c$ has a steep and asymmetric decrease (with respect to the parity-change points) towards the $\pi$ sector at the charge region (2,1). We support our interpretation with line-cuts taken from Fig.~\ref{fig1}e. Using fourth order perturbation we are able to obtain $\Gamma$ as to match the scale of $|I_c|$ to experiment \cite{NoteParameters} and find similar behavior in variations of gate for all line-cuts. 

In the vicinity of quadruple points, single dot states hybridize into bonding and anti-bonding molecular orbitals  \cite{van2002electron}.
We investigate the effect of inter-dot hybridization on $I_c$ via detuning, by which the dot levels are pulled away from each other \cite{van2002electron,su2017andreev}. Such line-cuts are shown in Figs.~\ref{fig5}a,b. These demonstrate that $I_c$ (white curve) can be tuned by changing the relative detuning between the two dot levels without changing the total charge of the quantum dots. At zero detuning (green arrows), at which inter-dot hybridization is highest, $I_c$ attains a local maximum. This occurs at a molecular doublet ground state $D_B$ of the form $\alpha\ket{D(0,1)}+\beta\ket{D(1,0)}$ in Fig.~\ref{fig5}a, and at a molecular singlet ground state $S_B$ of the form $\alpha\ket{S(1,1)}+\beta\ket{S(0,2)}$ in Fig.~\ref{fig5}b, where $\alpha$ and $\beta$ are coefficients. $I_c$ traces calculated by perturbation theory, shown in panels (c,d), are approximately consistent with the gate dependence and lineshape of $I_c$. However, due to the aforementioned cross coupling of $V_{gL}$ with the left tunnel barrier, the experiment does not display the exact symmetry with respect to zero detuning exhibited by the theory.

\begin{figure} [htb!]
\includegraphics[width=1\linewidth]{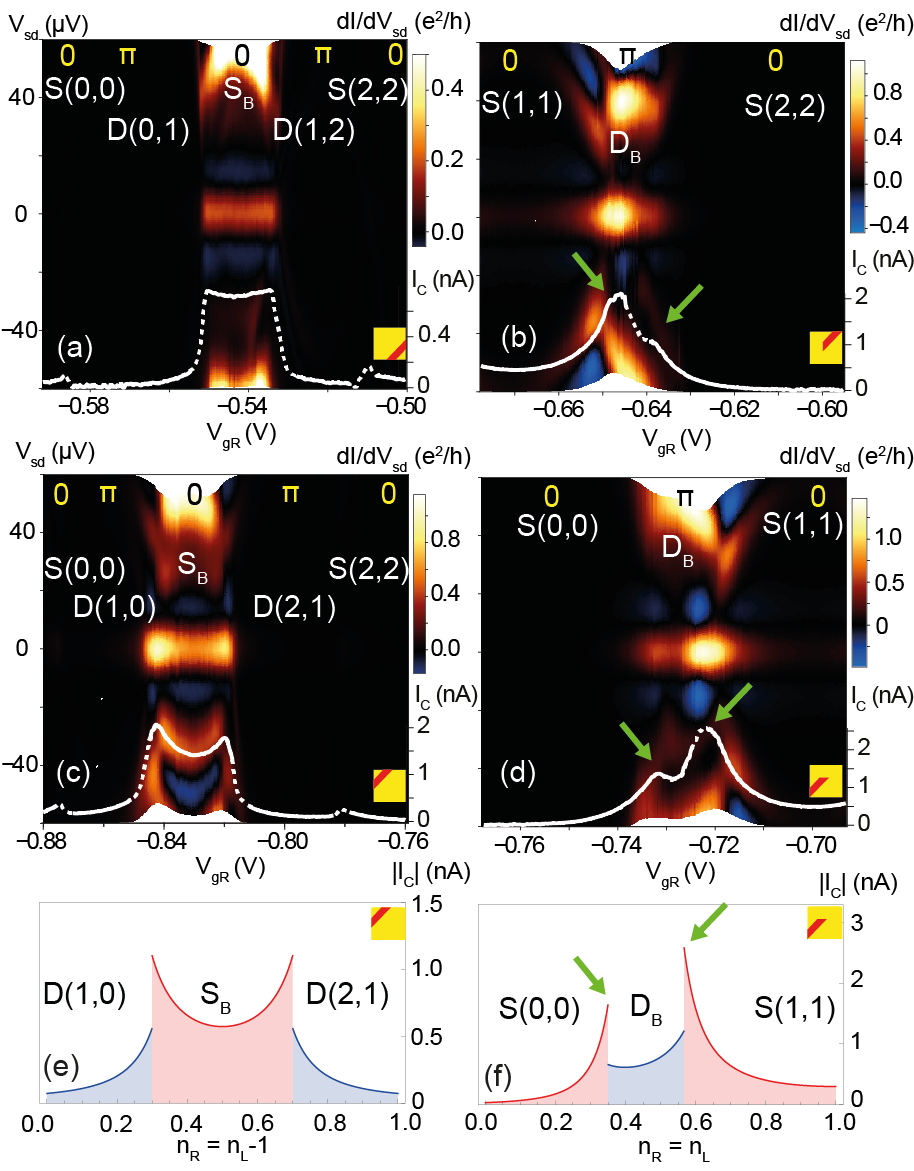}
\caption{(a-d) Line-cuts along the dotted lines in Fig. \ref{fig2}, through (a,c) $S_B$ and (b,d) $D_B$ molecular ground states. The levels are aligned per (a) $\epsilon_L=\epsilon_R+U_R$, (c) $\epsilon_L+U_L=\epsilon_R$, and (b,d) $\epsilon_L=\epsilon_R$. The difference is due to an additional electron in one of the levels in (a,c). \textit{Overlays}. Fitted $I_c$ (white line) \cite{NoteFit}. (e,f) Line-cuts of calculated $|I_c|$ through (e) $S_B$ and (f) $D_B$ ground states, done by fourth order perturbation theory. As before, red (blue) accounts for positive (negative) $I_c$. Note that line-cut (c) covers two additional charge sectors outside the range of the theory cut (e).}
\label{fig4}
\end{figure}

Next, we investigate the effect on $I_c$ of driving maximally-hybridized molecular orbitals (zero detuning) away from the Fermi level by taking line-cuts through the quadruple points following the dotted lines in Fig.~\ref{fig2}, so that the DQD energy levels are aligned and shifted simultaneously. Such line-cuts are shown in Figs.~\ref{fig4}a-d. Consistently with our interpretation of the diagram in Fig.~\ref{fig2}, $I_c$ is largest at the quadruple points, and quickly decreases in the Coulomb valleys. Interestingly, $I_c$ is symmetric with respect to the quadruple points at $V_{gR}=-0.54$ V in panel (a) and at $V_{gR}=-0.83$ V in panel (c), which correspond to line-cuts through a $S_B$ ground state, but it loses this symmetry in panels (b) and (d) with respect to $V_{gR}=-0.64$ V and $V_{gR}=-0.73$ V, respectively, which correspond to line-cuts through a $D_B$ ground state. In the two latter panels, $I_c$ is largest at the parity changes that go towards the (1,1) sector. This happens due to a greater availability of cotunneling channels in this sector, as captured by perturbation theory (see Figs.~\ref{fig4}e,f). Asymmetric $I_c$ peaks in panels (b,d) are indicated by green arrows. The observed symmetries stem from the combined effects of two types of symmetry. First, from the left-right symmetry $U_L\approx U_R$ as the system is then invariant under the transformation $(n_L,n_R)\rightarrow (n_R,n_L)$. Together with particle-hole symmetry $(n_L,n_R)\Rightarrow (2-n_L,2-n_R)$ this renders $I_c(n_L,n_R)=I_c(2-n_R,2-n_L)$ which corresponds to a mirror symmetry in the diagonal from $(2,0)$ to $(0,2)$. Line-cuts (a) and (c) are perpendicular to the diagonal and are therefore symmetrical around it while (b,d) are taken on each side of the diagonal and are therefore images of each other.

Ground state transitions observed in $I_c$ are consistent with crossings of Yu-Shiba-Rusinov (YSR) subgap states \cite{kirvsanskas2015yu,grove2017yu}. These can be observed in $dI/dV_{sd}(V_{sd})$ spectroscopy at larger $V_{sd}$ (see Supplemental Material). These states display a loop structure \cite{lee2014spin,jellinggaard2016tuning,deacon2010tunneling,pillet2013tunneling,lee2017scaling} and NDC \cite{kim2013transport,pillet2010andreev,grove2009superconductivity,eichler2007even,lee2012zero} similar to those in parity-changing single dots. Interestingly, simultaneously with the maximization of $I_c$ observed in detuning cuts in Figs.~\ref{fig5}a,b, the energy of the lowest pair of subgap states is minimized and displays an anti-crossing consistent with an absence of parity changes.

In this letter, we presented an experimental realization of a DQD Josephson junction, supported by the two impurity Anderson model. The stability diagram is electron-hole as well as left-right mirror symmetric. We showed that the discontinuities and gate-voltage asymmetries observed in the critical current are fully consistent with predictions of fourth-order perturbation theory. Finally, we probed molecular orbital ground states unique to the DQD system, finding a variation of $I_c$ with detuning from the level degeneracy.

\textbf{Acknowledgements}
\begin{acknowledgments}
We thank A. Jellinggaard, M.C. Hels, and  J. Ovesen for substrate preparation. We acknowledge financial support from the Carlsberg Foundation, the Independent Research Fund Denmark and the Danish National Research Foundation. P.~K. acknowledges support from Microsoft and the ERC starting grant no 716655 under the Horizon 2020 program. R.~\v{Z}. acknowledges support from the Slovenian Research Agency (ARRS) under P1-0044 and J1-7259.
\end{acknowledgments}


\begin{thebibliography}{36}
\bibitem[{Not({\natexlab{a}})}]{NoteParameters}
\bibinfo{note}{$t_d=0.3$ meV, $\Gamma_L=\Gamma_R=0.08$ meV -with
  $\Gamma_i=\pi\nu_F t^2_i$, $i=$L,R- were chosen to match the magnitude and
  gate dependence of $I_c$. For simplicity, $U_L=U_R=2$ meV, and $\Delta =
  0.25$ meV, which match closely the experimentally obtained values. For the
  same reason, the inter-dot charging energy was set to $U_{LR}=0$.}
\end{thebibliography}
\end{document}